\documentclass[aps,prl,twocolumn,showpacs,superscriptaddress,groupedaddress,floatfix]{revtex4}  
\usepackage{graphicx}  
\usepackage{dcolumn}   
\usepackage{bm}        
\usepackage{amsmath,amssymb}   
\usepackage[dvips]{color}
\usepackage{d0_style}
\usepackage{subfigure}
\usepackage{multirow}
\usepackage{afterpage}
\usepackage{url}

\def\Journal#1#2#3#4{{#1} {\bf #2}, #3 (#4)}

\def\NIMAs{Nucl. Instrum. Methods A}
\def\PLB{Phys.~Lett.~B}
\def\PRL{Phys.~Rev.~Lett.}
\def\PRD{Phys.~Rev.~D}
\def\PRDs{Phys. Rev. D}
\def\JHEP{J.~High Energy Phys. }

\newcommand{\AFBbb}{A^{b\bar{b}}_{\rm FB}}
\newcommand{\AFBtt}{A^{t\bar{t}}_{\rm FB}}
\newcommand{\BjpsiK}{B^{\pm} \rightarrow J/\psi K^{\pm}}

\newcommand{\Bjpsipi}{B^{\pm} \rightarrow J/\psi \pi^{\pm}}
\newcommand{\Bpm}{B^{\pm}}
\newcommand{\Kpm}{K^{\pm}}
\newcommand{\Jpsi}{J/\psi}
\newcommand{\phiKK}{\phi \rightarrow K^+K^-}
\newcommand{\sgn}{\mathop{\mathrm{sgn}}}
\newcommand{\beq}{\begin{equation}}
\newcommand{\eeq}{\end{equation}}

\begin{document}


\hspace{4.8in} \mbox{FERMILAB-PUB-14-472-E (accepted)}

\title{Measurement of the Forward-Backward Asymmetry in the Production of \boldmath$B^{\pm}$ Mesons in $p\bar{p}$ Collisions at $\sqrt{s}$ = 1.96 TeV}
\affiliation{LAFEX, Centro Brasileiro de Pesquisas F\'{i}sicas, Rio de Janeiro, Brazil}
\affiliation{Universidade do Estado do Rio de Janeiro, Rio de Janeiro, Brazil}
\affiliation{Universidade Federal do ABC, Santo Andr\'e, Brazil}
\affiliation{University of Science and Technology of China, Hefei, People's Republic of China}
\affiliation{Universidad de los Andes, Bogot\'a, Colombia}
\affiliation{Charles University, Faculty of Mathematics and Physics, Center for Particle Physics, Prague, Czech Republic}
\affiliation{Czech Technical University in Prague, Prague, Czech Republic}
\affiliation{Institute of Physics, Academy of Sciences of the Czech Republic, Prague, Czech Republic}
\affiliation{Universidad San Francisco de Quito, Quito, Ecuador}
\affiliation{LPC, Universit\'e Blaise Pascal, CNRS/IN2P3, Clermont, France}
\affiliation{LPSC, Universit\'e Joseph Fourier Grenoble 1, CNRS/IN2P3, Institut National Polytechnique de Grenoble, Grenoble, France}
\affiliation{CPPM, Aix-Marseille Universit\'e, CNRS/IN2P3, Marseille, France}
\affiliation{LAL, Universit\'e Paris-Sud, CNRS/IN2P3, Orsay, France}
\affiliation{LPNHE, Universit\'es Paris VI and VII, CNRS/IN2P3, Paris, France}
\affiliation{CEA, Irfu, SPP, Saclay, France}
\affiliation{IPHC, Universit\'e de Strasbourg, CNRS/IN2P3, Strasbourg, France}
\affiliation{IPNL, Universit\'e Lyon 1, CNRS/IN2P3, Villeurbanne, France and Universit\'e de Lyon, Lyon, France}
\affiliation{III. Physikalisches Institut A, RWTH Aachen University, Aachen, Germany}
\affiliation{Physikalisches Institut, Universit\"at Freiburg, Freiburg, Germany}
\affiliation{II. Physikalisches Institut, Georg-August-Universit\"at G\"ottingen, G\"ottingen, Germany}
\affiliation{Institut f\"ur Physik, Universit\"at Mainz, Mainz, Germany}
\affiliation{Ludwig-Maximilians-Universit\"at M\"unchen, M\"unchen, Germany}
\affiliation{Panjab University, Chandigarh, India}
\affiliation{Delhi University, Delhi, India}
\affiliation{Tata Institute of Fundamental Research, Mumbai, India}
\affiliation{University College Dublin, Dublin, Ireland}
\affiliation{Korea Detector Laboratory, Korea University, Seoul, Korea}
\affiliation{CINVESTAV, Mexico City, Mexico}
\affiliation{Nikhef, Science Park, Amsterdam, the Netherlands}
\affiliation{Radboud University Nijmegen, Nijmegen, the Netherlands}
\affiliation{Joint Institute for Nuclear Research, Dubna, Russia}
\affiliation{Institute for Theoretical and Experimental Physics, Moscow, Russia}
\affiliation{Moscow State University, Moscow, Russia}
\affiliation{Institute for High Energy Physics, Protvino, Russia}
\affiliation{Petersburg Nuclear Physics Institute, St. Petersburg, Russia}
\affiliation{Instituci\'{o} Catalana de Recerca i Estudis Avan\c{c}ats (ICREA) and Institut de F\'{i}sica d'Altes Energies (IFAE), Barcelona, Spain}
\affiliation{Uppsala University, Uppsala, Sweden}
\affiliation{Taras Shevchenko National University of Kyiv, Kiev, Ukraine}
\affiliation{Lancaster University, Lancaster LA1 4YB, United Kingdom}
\affiliation{Imperial College London, London SW7 2AZ, United Kingdom}
\affiliation{The University of Manchester, Manchester M13 9PL, United Kingdom}
\affiliation{University of Arizona, Tucson, Arizona 85721, USA}
\affiliation{University of California Riverside, Riverside, California 92521, USA}
\affiliation{Florida State University, Tallahassee, Florida 32306, USA}
\affiliation{Fermi National Accelerator Laboratory, Batavia, Illinois 60510, USA}
\affiliation{University of Illinois at Chicago, Chicago, Illinois 60607, USA}
\affiliation{Northern Illinois University, DeKalb, Illinois 60115, USA}
\affiliation{Northwestern University, Evanston, Illinois 60208, USA}
\affiliation{Indiana University, Bloomington, Indiana 47405, USA}
\affiliation{Purdue University Calumet, Hammond, Indiana 46323, USA}
\affiliation{University of Notre Dame, Notre Dame, Indiana 46556, USA}
\affiliation{Iowa State University, Ames, Iowa 50011, USA}
\affiliation{University of Kansas, Lawrence, Kansas 66045, USA}
\affiliation{Louisiana Tech University, Ruston, Louisiana 71272, USA}
\affiliation{Northeastern University, Boston, Massachusetts 02115, USA}
\affiliation{University of Michigan, Ann Arbor, Michigan 48109, USA}
\affiliation{Michigan State University, East Lansing, Michigan 48824, USA}
\affiliation{University of Mississippi, University, Mississippi 38677, USA}
\affiliation{University of Nebraska, Lincoln, Nebraska 68588, USA}
\affiliation{Rutgers University, Piscataway, New Jersey 08855, USA}
\affiliation{Princeton University, Princeton, New Jersey 08544, USA}
\affiliation{State University of New York, Buffalo, New York 14260, USA}
\affiliation{University of Rochester, Rochester, New York 14627, USA}
\affiliation{State University of New York, Stony Brook, New York 11794, USA}
\affiliation{Brookhaven National Laboratory, Upton, New York 11973, USA}
\affiliation{Langston University, Langston, Oklahoma 73050, USA}
\affiliation{University of Oklahoma, Norman, Oklahoma 73019, USA}
\affiliation{Oklahoma State University, Stillwater, Oklahoma 74078, USA}
\affiliation{Brown University, Providence, Rhode Island 02912, USA}
\affiliation{University of Texas, Arlington, Texas 76019, USA}
\affiliation{Southern Methodist University, Dallas, Texas 75275, USA}
\affiliation{Rice University, Houston, Texas 77005, USA}
\affiliation{University of Virginia, Charlottesville, Virginia 22904, USA}
\affiliation{University of Washington, Seattle, Washington 98195, USA}
\author{V.M.~Abazov} \affiliation{Joint Institute for Nuclear Research, Dubna, Russia}
\author{B.~Abbott} \affiliation{University of Oklahoma, Norman, Oklahoma 73019, USA}
\author{B.S.~Acharya} \affiliation{Tata Institute of Fundamental Research, Mumbai, India}
\author{M.~Adams} \affiliation{University of Illinois at Chicago, Chicago, Illinois 60607, USA}
\author{T.~Adams} \affiliation{Florida State University, Tallahassee, Florida 32306, USA}
\author{J.P.~Agnew} \affiliation{The University of Manchester, Manchester M13 9PL, United Kingdom}
\author{G.D.~Alexeev} \affiliation{Joint Institute for Nuclear Research, Dubna, Russia}
\author{G.~Alkhazov} \affiliation{Petersburg Nuclear Physics Institute, St. Petersburg, Russia}
\author{A.~Alton$^{a}$} \affiliation{University of Michigan, Ann Arbor, Michigan 48109, USA}
\author{A.~Askew} \affiliation{Florida State University, Tallahassee, Florida 32306, USA}
% ...
%
%
\collaboration{The D0 Collaboration\footnote{with visitors from
$^{a}$Augustana College, Sioux Falls, SD, USA,
$^{b}$The University of Liverpool, Liverpool, UK,
$^{c}$DESY, Hamburg, Germany,
$^{d}$Universidad Michoacana de San Nicolas de Hidalgo, Morelia, Mexico
$^{e}$SLAC, Menlo Park, CA, USA,
$^{f}$University College London, London, UK,
$^{g}$Centro de Investigacion en Computacion - IPN, Mexico City, Mexico,
$^{h}$Universidade Estadual Paulista, S\~ao Paulo, Brazil,
$^{i}$Karlsruher Institut f\"ur Technologie (KIT) - Steinbuch Centre for Computing (SCC),
D-76128 Karlsruhe, Germany,
$^{j}$Office of Science, U.S. Department of Energy, Washington, D.C. 20585, USA,
$^{k}$American Association for the Advancement of Science, Washington, D.C. 20005, USA,
$^{l}$Kiev Institute for Nuclear Research, Kiev, Ukraine,
$^{m}$University of Maryland, College Park, Maryland 20742, USA
and
$^{n}$European Orgnaization for Nuclear Research (CERN), Geneva, Switzerland
}} \noaffiliation
\vskip 0.25cm
\date{December 22, 2014}

\begin{abstract} We present a measurement of the forward-backward asymmetry in the production of $\Bpm$ mesons, $A_{\rm FB}(\Bpm)$,
using $B^{\pm} \rightarrow J/\psi K^{\pm}$ decays in 10.4 ${\rm fb}^{-1}$ of $p\bar{p}$ collisions at $\sqrt{s} = 1.96$ TeV collected by the D0 
experiment during Run II of the Tevatron collider. A nonzero asymmetry would indicate a preference for a particular flavor, i.e., $b$ quark
or $\bar{b}$ antiquark, to be produced in the direction of the proton beam. We extract $A_{\rm FB}(\Bpm)$ from a maximum likelihood fit to the 
difference between the numbers of forward- and backward-produced $B^{\pm}$ mesons. We measure an asymmetry consistent with zero: $A_{\rm FB}(\Bpm)$ = 
[$-$0.24 $\pm$ 0.41\,(stat)\,$\pm$ 0.19\,(syst)]\%. 
\end{abstract}

\pacs{13.25.Hw, 11.30.Er}
\maketitle

Over the few past years there has been much interest in the forward-backward asymmetry in $t\bar{t}$ production ($\AFBtt$)~\cite{theory2}, 
especially since initial experimental results were larger than standard model (SM) predictions~\cite{ttDzero1,ttCDF}. 
These observations prompted development of models beyond the SM that could explain the excess~\cite{ttmodels}. 
The corresponding asymmetry in $b\bar{b}$ production, $\AFBbb$, has the same sources as $\AFBtt$ but is expected to have a smaller 
magnitude in the SM, making it an important probe of these new physics models~\cite{Murphy,Ipek}. 

The most recent D0 measurements of $\AFBtt$~\cite{ttDzero} agree with the SM~\cite{Mitov}. A closely related quantity called the $t\bar{t}$ charge 
asymmetry has been studied at the LHC~\cite{ttCMS,ttAtlas}. The LHCb collaboration has recently measured the charge asymmetry between $b$ and 
$\bar{b}$ jets in $pp$ collisions~\cite{lhcb}.

A forward-backward asymmetry in the production of heavy quark $Q$ is primarily caused by interference between tree-level and loop diagrams for 
$q\bar{q} \rightarrow Q\bar{Q}$ interactions, and also by interference between initial and final state gluon radiation~\cite{KuhnRodrigo}.  
We measure the forward-backward asymmetry using fully reconstructed $\Bpm \rightarrow \Jpsi(\rightarrow \mu^+\mu^-)\Kpm$ decays where the 
$\Bpm$ directly identifies the quark flavor (i.e., $b$ or $\bar{b}$). Compared to $b$ jet reconstruction, this method has the advantages that 
the charge of the $b$ quark is unambiguously determined, and there is no need to account for $B^0/\bar{B^0}$ oscillations.
The quantity $A_{\rm FB}(\Bpm)$ is sensitive to the same production asymmetries as $\AFBbb$. In $p\bar{p}$ collisions, the forward category 
indicates a $b$\,($\bar{b}$) quark, or $B^-$\,($B^+$) meson, emitted with a longitudinal momentum component in the direction of the proton\,
(antiproton) beam. 

We reconstruct a $\Bpm$ meson and categorize it as forward or backward with a variable $q_{\rm FB} = -q_B\sgn(\eta_B)$, where $q_B$ is the $\Bpm$ 
meson electric charge, $\sgn(x)$ is the sign function, and $\eta_B$ is the $\Bpm$ meson pseudorapidity~\cite{note1}. The forward-backward asymmetry 
of the $\Bpm$ mesons is
\beq\label{eq:afb}
A_{\rm FB}(\Bpm) = \frac{N(q_{\rm FB} > 0) - N(q_{\rm FB} < 0)}{N(q_{\rm FB} > 0) + N(q_{\rm FB} < 0)}.
\eeq

Inclusive predictions of $\AFBbb$ give positive asymmetries of $\approx0.5$\%~\cite{Murphy,Manohar}, but
the mass scales of the $b\bar{b}$ pairs considered [$M(b\bar{b}) > 35$ GeV, or $p(b) >$ $\approx15$ GeV] are more relevant 
for a jet-based analysis. To make SM predictions tailored to our kinematics and selections, 
we produce next-to-leading-order Monte Carlo (MC) samples for QCD production of $\Bpm$ in the process $p\bar{p} \rightarrow b\bar{b} X$.  
MC events are generated using {\sc mc@nlo}~\cite{MCNLO} with parton distribution function (PDF) set {\sc cteq6m1}~\cite{cteq} 
and HERWIG~\cite{herwig} for parton showering and hadronization. Detector simulation is performed using {\sc geant3}~\cite{geant}. 

The D0 experiment collected data at $\sqrt{s}$ = 1.96 TeV during Run II of the Fermilab Tevatron $p\bar{p}$ collider, from 2002 
until the Tevatron shutdown in 2011. The D0 detector is described in detail elsewhere~\cite{dzero}. For this analysis 
the most important detector elements are the central tracking and muon systems. The central tracking system consists of a 
silicon microstrip tracker and a central fiber tracker, both located within a 1.9~T superconducting solenoidal magnet, with 
designs optimized for tracking and vertex finding at pseudorapidities $|\eta| < 3$ and $|\eta| < 2.5$, respectively.
The muon system has a layer of tracking detectors and scintillation trigger counters outside a liquid argon sampling calorimeter 
and two similar layers outside a 1.8~T iron toroid~\cite{muonNIM}, and covers the region $|\eta_{\rm det}| \approx 2$ where 
$|\eta_{\rm det}|$ is measured from the center of the detector.
The solenoid and toroid magnet polarities were reversed approximately every two weeks giving nearly equal beam exposure 
to each polarity combination. The data used in this analysis were collected with a suite of single muon and dimuon triggers.

We select $\BjpsiK$ candidates from the D0 Run II data set with an integrated luminosity of 10.4 ${\rm fb}^{-1}$. 
Candidates are reconstructed by identifying a pair of oppositely charged muons (decay products of the $\Jpsi$ meson) produced 
along with a charged track (the $\Kpm$ candidate) at a common vertex displaced from the $p\bar{p}$ interaction vertex.

All tracks must lie within the pseudorapidity coverage of the muon and central tracking systems, 
$\lvert\eta\rvert < 2.1$. Selected muons have transverse momentum $p_T > 1.5$ GeV, and $\Kpm$ candidates have $p_T > 0.7$ 
GeV. At least one muon must traverse both inner and outer layers of the muon detector. Both muons must match to tracks 
in the central tracking system. The $\Jpsi$ candidates with reconstructed invariant mass $M(\mu^+\mu^-)$ between 2.7 and 3.45 GeV are accepted 
if their transverse decay length ($L_{xy}$) uncertainty is less than 0.1 cm, where $L_{xy}$ is the distance from the $p\bar{p}$ 
vertex to a particle's decay vertex in the $x$-$y$ plane. The cosine of the pointing angle~\cite{note2} must be greater than zero. 

The combination of $\mu^+$, $\mu^-$, and $\Kpm$ tracks to form a $\Bpm$ decay vertex must have $\chi^2 < 16$ for 3 degrees of freedom, 
and the cosine of the $\Bpm$ pointing angle must be above 0.8. $\Bpm$ candidates are accepted if they are significantly displaced from the 
$p\bar{p}$ vertex. Their transverse decay length significance (defined as $L_{xy}$ divided by its uncertainty) must be greater than three. To 
calculate the $\Bpm$ candidate mass we correct the muon momenta by constraining $M(\mu^+\mu^-)$ to the world average $\Jpsi$ meson 
mass~\cite{Jmass}. The selected $\Bpm$ mass range is 5.05 -- 5.65 GeV. 

Because definitions of forward and backward are tied directly to $\sgn(\eta_B)$, the ambiguous region near $\eta_B = 0$ is given special 
consideration. We compare $\eta$ of the $\Bpm$ mesons and their parent $b$ quarks at the production and reconstruction levels in {\sc mc@nlo}. 
Rejecting events with $\lvert\eta_B\rvert < 0.1$ removes all $\Bpm$ mesons reconstructed with incorrect $q_{\rm FB}$ without significantly affecting
$A_{\rm FB}(\Bpm)$. After the cut, more than 99.9\% of $\Bpm$ mesons give the same $q_{\rm FB}$ as the parent $b$ quark, indicating minimal 
hadronization effects on $A_{\rm FB}(\Bpm)$. The distribution of ($\eta_b - \eta_B$) has a rms width of 0.11.

Background rejection is improved using a boosted decision tree (BDT)~\cite{BDT} trained on a simulated MC signal sample and a background sample from data sidebands 
around the selected $\Bpm$ mass range (4.0 -- 5.05 and 5.65 -- 7.0 GeV). Leading-order signal MC events are generated with {\sc pythia}~
\cite{pythia} and processed through the same reconstruction code used for data. We weight MC events so that the $p_T$ distributions of the muons 
match the distributions in data, which are affected by trigger inefficiencies. Additional weights are applied to match distributions of 
$p_T(\Bpm)$, $p_T(\Kpm)$, and $\chi^2$ of the $\Bpm$ decay vertex fit to data distributions. Finally, we weight MC events so that the probability 
of reconstructing isolated muons or $\Bpm$ candidates matches the probability measured in data. Isolated particles have no other tracks in a cone 
of size $\Delta\mathcal{R} = 1$ around them, where $\Delta\mathcal{R} = \sqrt{\Delta\phi^2 + \Delta\eta^2}$ is the angular separation between 
tracks. This weighting gives optimal agreement between data and simulation in all 40 BDT input variables, which include particle momenta, distances 
from the $p\bar{p}$ vertex, decay lengths, pointing angles, isolation of the muons and $\Bpm$ meson, and azimuthal angle separation for various 
particle pairs. A cut on the BDT discriminant is chosen to minimize the statistical uncertainty of $A_{\rm FB}(\Bpm)$. After all cuts we find one 
$\Bpm$ candidate in 98.5\% of events, with the remainder having two or more candidates. All candidates are used independently in this analysis.

We extract $A_{\rm FB}(\Bpm)$ from a maximum likelihood fit incorporating a signal probability distribution and three background distributions (see 
below), which are functions of the reconstructed $\Bpm$ mass $m_{\Jpsi K}$ and the kaon energy $E_K$. The signal distribution $S(m_{\Jpsi K}, E_K)$ is 
modeled by a double-Gaussian function with six parameters, where both Gaussians have the same mean but different widths. The widths have an 
exponential dependence on $E_K$. Signal parameters are allowed to differ for the $\eta < -0.1$ and $\eta > 0.1$ regions to account for slight 
differences in the magnetic field along the beam direction.

The background distribution $P(m_{\Jpsi K}, E_K)$ describes $\Bjpsipi$ events where the pion is assigned the kaon mass, creating an artificially high 
reconstructed $\Bpm$ mass. Distribution $P$ is a reflection of $S$ with the mean mass value shifted to account for the $K/\pi$ mass difference and 
the widths scaled by a ratio of the mean mass values. Background distribution $T(m_{\Jpsi K})$ describes partially reconstructed decays of type $B_x 
\rightarrow \Jpsi h^{\pm}X$, which have reconstructed mass lower than the $\Bpm$ mass. Distribution $T$ is empirically modeled using a threshold 
function with a floating inflection point and the slope fixed from MC simulation~\cite{iainNote,bsmumu}. Finally, the background distribution 
$E(m_{\Jpsi K}, E_K)$ describes combinatoric background and is modeled using an exponential function with three parameters, where the slope depends 
on $E_K$.

The unbinned fit minimizes LLH, the negative log of the likelihood function $\mathcal{L}_n$ summed over $N$ selected $\Bpm$ candidates, each with 
weight $w_n$ (defined below): 
\beq 
{\rm LLH} = -2\sum_{n=1}^N w_n \ln(\mathcal{L}_n). 
\eeq 
Here $\mathcal{L}_n$ is a function of the four probability density distributions, with each assigned sample fraction $f_i$ and forward-backward 
asymmetry $A_i$. While systematic effects were studied, the $A_i$ parameters were blinded by adding unknown random offsets. 
The likelihood $\mathcal{L}_n$ has 26 parameters and is normalized to 1: 
\begin{align}\label{eq:PDF} 
\mathcal{L}_n = &\:\alpha(E_K)[f_S(1 + q_{\rm FB}A_S)S+ f_P(1 + q_{\rm FB}A_P)P\nonumber\\ 
              &+ f_T(1 + q_{\rm FB}A_T)T] + f_E(1 + q_{\rm FB}A_E)E, 
\end{align}
where $f_E$ = $[1 - \alpha(E_K)(f_S + f_P + f_T)]$ and $\alpha(E_K)$ uses three parameters to describe the dependence of the sample fractions on 
$E_K$~\cite{iainNote}.

Asymmetries in the detector material and $\Jpsi$ or $\Kpm$ reconstruction between $\eta < 0$ (the ``north'' side of the detector) and $\eta > 0$  
(the ``south'' side) can result in an apparent $A_{\rm FB}$. A north-south asymmetry is defined as $A_{\rm NS} = (N_N - N_S)/(N_N + N_S)$.
Because $B^+$ and $B^-$ particles on the same side of the detector have opposite $q_{\rm FB}$, corrections for north-south efficiency differences 
will generally cancel when determining $A_{\rm FB}(\Bpm)$. We measure $A_{\rm NS}$ in data samples with no expected production asymmetries. 
Decays of $\phiKK$ are used to measure $A_{\rm NS}(\Kpm)$. Signal and background models are determined from MC simulation and a $\chi^2$ 
minimization fit is performed simultaneously on north- and south-side data. We measure $A_{\rm NS}(\Kpm)$ in bins of leading 
kaon $\lvert\eta\rvert$; there is no significant dependence on $p_T$. Integrated over all $\lvert\eta\rvert$, $A_{\rm NS}(K^+)$ = 
(0.39 $\pm$ 0.22)\% and $A_{\rm NS}(K^-)$ = (0.64 $\pm$ 0.23)\%.

We measure $A_{\rm NS}(\Jpsi)$ using prompt $\Jpsi \rightarrow \mu^+\mu^-$ decays. $\Jpsi$ mesons with significant $L_{xy}$ are generally from $B$ 
decays which could exhibit a north-south asymmetry due to $A_{\rm FB}(\Bpm)$. To reduce the fraction of nonprompt $\Jpsi$ mesons to a negligible 
level we require the $\Jpsi$ $L_{xy}$ significance to be less than 1.5. Background events under the peak from 2.9 -- 3.3 GeV are removed with a 
sideband subtraction, and $A_{\rm NS}(\Jpsi)$ is calculated in bins of $\vert\eta\rvert$ and $p_T$. Integrated over all $\lvert\eta\rvert$ and $p_T$, 
$A_{\rm NS}(\Jpsi)$ = ($-$0.41 $\pm$ 0.04)\%. 

Measured $A_{\rm NS}$ values are used to determine ``efficiency weights'' $w_{\Kpm}$ and $w_{\Jpsi}$ that equalize the relative reconstruction 
efficiencies on both sides of the detector. Applying these weights has a small effect on $A_{\rm FB}(\Bpm)$: a shift of 0.06\% from $w_{\Kpm}$ and a 
shift of $-$0.01\% from $w_{\Jpsi}$. Uncertainties on $A_{\rm NS}(\Jpsi)$ and $A_{\rm NS}(\Kpm)$ contribute an uncertainty of 0.003\% to 
$A_{\rm FB}(\Bpm)$, determined using an ensemble test with 500 Gaussian variations of the $A_{\rm NS}$ values. 

The total event weight is $w_n = w_{\rm magnet}w_{\Kpm}w_{\Jpsi}$, where $w_{\rm magnet}$ equalizes the number of events in eight settings of 
solenoid polarity, toroid polarity, and $\Bpm$ charge. Equalizing the contribution from each magnet polarity combination removes tracking charge 
asymmetries to first order, since in one polarity a $B^+$ is reconstructed with the same sign of curvature as a $B^-$ in the opposite polarity. 
Also equalizing the number of $B^+$ and $B^-$ candidates eliminates the need to correct for different $K^+$ and $K^-$ interaction cross sections in 
the detector~\cite{dimuACP}.

\begin{figure}[htbp] 
\centering 
\includegraphics[width=3.25in]{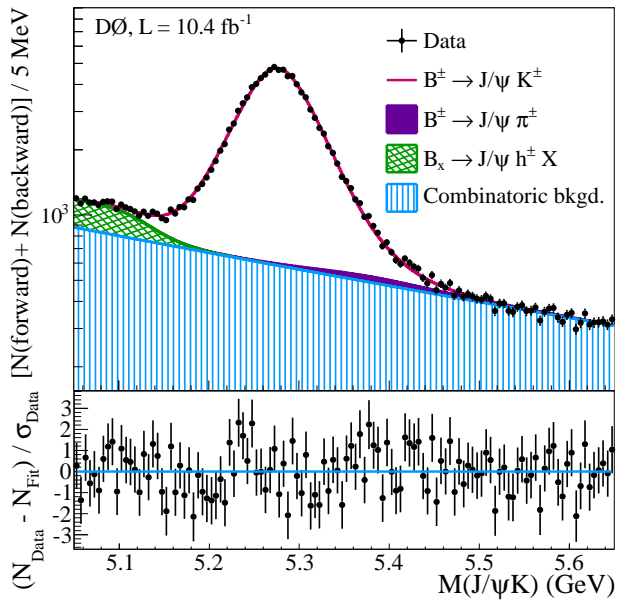} 
\caption{(color online) Invariant mass $M(\Jpsi K)$ of (forward + backward) events with fitted distributions. The lower pane shows the residuals.} 
\label{fig:F+B} 
\end{figure}

\begin{figure}[hbtp]
\centering
\includegraphics[width=3.25in]{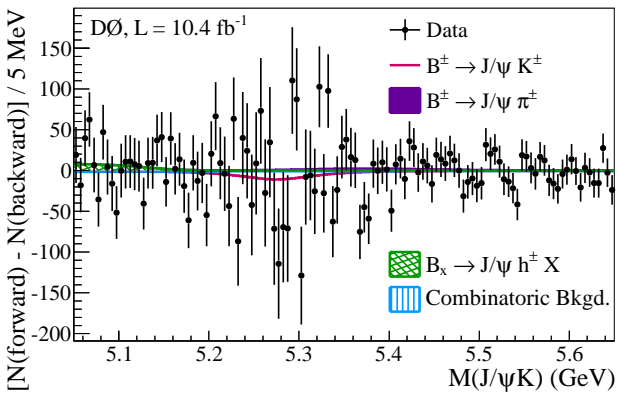}
\caption{(color online) Invariant mass $M(\Jpsi K)$ of (forward $-$ backward) events with fitted distributions which include the asymmetry 
parameters $A_i$.}
\label{fig:F-B}
\end{figure}

The weighted data sample contains 160\,360 $\Bpm$ candidates and the fit yields 89\,328 $\pm$ 349 $\BjpsiK$ decays. Although the fit was unbinned, 
to visualize the data and fit quality, binned distributions of invariant mass $M(\Jpsi K)$ for the sum and the difference in the numbers of forward and 
backward $B^{\pm}$ candidates with their projected fits are shown in Figs.~\ref{fig:F+B} and~\ref{fig:F-B}. Over both mass 
distributions we obtain $\chi^2$/ndf = 249/214.
We measure a signal asymmetry consistent with zero: $A_{\rm FB}(\Bpm)$ = [$-$0.24 $\pm$ 0.41\,(stat)]\%. The asymmetry is consistent over time and 
with $B^+$ and $B^-$ samples fitted separately. Asymmetries of the background distributions are also consistent with zero.

To determine systematic uncertainties on $A_{\rm FB}(\Bpm)$ a number of variations are made to the analysis. 
Data sample variations include training four alternative BDTs with different variables or input samples and using a range of BDT discriminant cuts. 
Fit variations include varying the $\Bpm$ mass range, removing dependences on $E_K$ from the distributions, 
allowing the slope of $T(m_{\Jpsi K})$ to float, and fixing the background asymmetry parameters to zero.

To estimate the systematic error from the reconstruction asymmetries we measure $A_{\rm NS}(\Jpsi)$ and $A_{\rm NS}(\Kpm)$ using alternate data samples and 
calculations in different bins or with alternate fit parameters. Biases in the fitting procedure are explored with ensemble tests on randomized 
data, comparing input and fitted values of $A_{\rm FB}(\Bpm)$. No bias is observed, and a systematic uncertainty is assigned based on the spread of 
results in the ensemble test.
The total systematic uncertainty on the data measurement is 0.19\%, as summarized in Table~\ref{tab:systuncerts}. 
\begin{table}[htbp]
\caption{Summary of uncertainties on $A_{\rm FB}(\Bpm)$ in data.}\label{tab:systuncerts}
\centering
\begin{ruledtabular}\begin{tabular}{lc}
Source       & Uncertainty \\
\hline
Statistical & 0.41\% \\
\hline
Alternative BDTs and cuts & 0.17\% \\
Fit variations & 0.06\% \\
Reconstruction asymmetries & 0.05\% \\
Fit bias & 0.02\% \\
\hline
Systematic uncertainty & 0.19\% \\
\hline
Total uncertainty & 0.45\% \\
\end{tabular}\end{ruledtabular}
\end{table}

To compare this measurement to the SM, the {\sc mc@nlo} simulation is analyzed as described above, applying $\BjpsiK$ selections and weights to 
correct for muon trigger effects. Additionally, reconstructed muon and kaon tracks must match tracks from generated $\BjpsiK$ decays. Since 
matching reconstructed and generated $\Bpm$ mesons leaves no background events, $A_{\rm FB}^{\rm SM}(\Bpm)$ is calculated directly according to 
Eq.~(\ref{eq:afb}). 

The dominant systematic uncertainty on $A_{\rm FB}^{\rm SM}(\Bpm)$ is due to renormalization and factorization energy scale choices. {\sc mc@nlo} 
defines $\mu_R$ and $\mu_F$ for renormalization and factorization energy scales~\cite{MCNLO} as the square root of the average of $m^2_T = m^2 + 
p_T^2$ for the $b$ and $\bar{b}$ quarks~\cite{SFemail}, with $b$ quark mass $m$ set to 4.75 GeV. Since $\AFBbb$ is zero at leading order, there is 
a large scale dependence in predictions at next-to-leading order~\cite{ellis}. Both scales are varied independently from $\frac{1}{2}\mu_{R,F}$ 
to $2\mu_{R,F}$ to estimate an uncertainty due to uncalculated higher orders. Half the largest spread of variations gives a systematic uncertainty 
of 0.44\%. The uncertainty on $A_{\rm FB}^{\rm SM}(\Bpm)$ due to $b$ quark fragmentation is estimated by weighting events so the distribution of 
$p(\Bpm)_{||}/p(b)$ matches a Bowler function~\cite{frag} tuned to LEP data or SLD data, where $p(\Bpm)_{||}$ is the component of $p(\Bpm)$ in the 
$b$ quark direction. Half the largest spread of variations to $A_{\rm FB}^{\rm SM}(\Bpm)$ is 0.25\%. The negligible PDF uncertainty of 0.03\% is 
calculated by varying the twenty {\sc cteq6m1} eigenvectors by their uncertainties and determining the standard deviation of the variations. We 
find $A_{\rm FB}^{\rm SM}(\Bpm)$ = [2.31 $\pm$ 0.34\,(stat) $\pm$ 0.51\,(syst)]\%. Combining all data and MC uncertainties in quadrature, the 
{\sc mc@nlo} result differs from data by (2.55 $\pm$ 0.76)\%, or 3.3 standard deviations. 

\begin{figure}[htbp]\begin{center}
\includegraphics[width=3.25in]{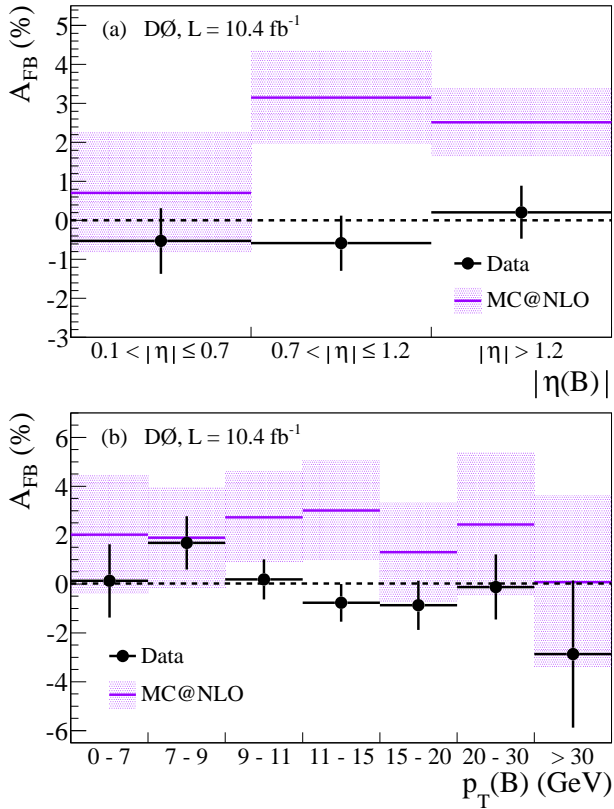}
\caption{(color online) Comparison of $A_{\rm FB}(\Bpm)$ and $A_{\rm FB}^{\rm SM}(\Bpm)$ in bins of (a) $\lvert\eta_B\rvert$ and (b) $p_T(B)$. Data 
points and MC bands include statistical uncertainties convoluted with systematic uncertainties.}
\label{fig:mccomps}
\end{center}\end{figure}

Figure~\ref{fig:mccomps} shows measurements of $A_{\rm FB}(\Bpm)$ and $A_{\rm FB}^{\rm SM}(\Bpm)$ versus transverse momentum and pseudorapidity. 
The fully reconstructed $\Jpsi\Kpm$ final state produces good kinematic agreement between reconstructed and generated $\Bpm$ mesons, so corrections 
to recover the true $\Bpm$ kinematics are unnecessary. The average $p_T$ of the $\Bpm$ mesons is 12.9 GeV. We find that $A_{\rm FB}(\Bpm)$ is 
systematically lower than $A_{\rm FB}^{\rm SM}(\Bpm)$ for all pseudorapidities, and for $p_T(B)$ = 9 -- 30 GeV. Considering the MC systematic 
uncertainties to be correlated\,(uncorrelated), Fig.~\ref{fig:mccomps}\,(a) has $\chi^2$ = 10.3\,(11.8) for three bins and Fig.~\ref{fig:mccomps}\,
(b) has $\chi^2$ = 6.6\,(7.0) for seven bins. 

In conclusion, we have measured the forward-backward asymmetry in the production of $\Bpm$ mesons with $\BjpsiK$ decays in $p\bar{p}$ collisions 
at $\sqrt{s} = 1.96$ TeV. For $\Bpm$ mesons with a mean $p_T$ of 12.9 GeV, the result is $A_{\rm FB}(\Bpm)$ = [$-$0.24 $\pm$ 0.41\,(stat) $\pm$ 0.19\,
(syst)]\%, which is the first measurement of this quantity. The observed discrepancy of $\approx3$ standard deviations between our measurement and 
the {\sc mc@nlo} estimate suggests that more rigorous determination of the standard model prediction is needed to interpret these results.

%

We thank the staffs at Fermilab and collaborating institutions,
and acknowledge support from the
Department of Energy and National Science Foundation (United States of America);
Alternative Energies and Atomic Energy Commission and
National Center for Scientific Research/National Institute of Nuclear and Particle Physics  (France);
Ministry of Education and Science of the Russian Federation, 
National Research Center ``Kurchatov Institute" of the Russian Federation, and 
Russian Foundation for Basic Research  (Russia);
National Council for the Development of Science and Technology and
Carlos Chagas Filho Foundation for the Support of Research in the State of Rio de Janeiro (Brazil);
Department of Atomic Energy and Department of Science and Technology (India);
Administrative Department of Science, Technology and Innovation (Colombia);
National Council of Science and Technology (Mexico);
National Research Foundation of Korea (Korea);
Foundation for Fundamental Research on Matter (The Netherlands);
Science and Technology Facilities Council and The Royal Society (United Kingdom);
Ministry of Education, Youth and Sports (Czech Republic);
Bundesministerium f\"{u}r Bildung und Forschung (Federal Ministry of Education and Research) and 
Deutsche Forschungsgemeinschaft (German Research Foundation) (Germany);
Science Foundation Ireland (Ireland);
Swedish Research Council (Sweden);
China Academy of Sciences and National Natural Science Foundation of China (China);
and
Ministry of Education and Science of Ukraine (Ukraine).


\begin{thebibliography}{99}
\bibitem{note1}The D0 collaboration defines a coordinate system with the $z$ axis along the proton beam direction, 
the $x$ axis pointing away from the Tevatron center, and the $y$ axis pointing upwards. Pseudorapidity is defined 
as $\eta = -\ln[\tan(\theta/2)]$, where $\theta$ is the laboratory frame polar angle. Angle $\phi$ is the azimuthal angle 
in the $x$-$y$ plane.
\bibitem{note2}The pointing angle is defined as the angle between a particle's momentum vector and the vector from the $p\bar{p}$ vertex to the particle's decay vertex,
with vectors defined in the $x$-$y$ plane.
\end{thebibliography}
\end{document}